**A Monte Carlo simulation of a protein (CoVE) in a matrix of random barriers**

R.B. Pandey
School of Mathematics and Natural Sciences, University of Southern Mississippi, Hattiesburg, MS 39406-5043, USA
Email: ras.pandey@usm.edu

**Abstract:** Monte Carlo simulations are performed to study structure and dynamics of a protein CoVE in random media generated by a random distribution of barriers at concentration $c$ with a coarse-grained model in its native (low temperature) and denatured (high temperature) phase. The stochastic dynamics of the protein is diffusive in denature phase at low $c$, it slows down on increasing $c$ and stops moving beyond a threshold ($c_{th} \approx 0.10$). In native phase, the protein moves extremely slow at low $c$ but speeds up on further increasing $c$ in a characteristic range ($c = 0.10 - 0.20$) before getting trapped at high $c$ ($c_{th} \approx 0.30$). The radius of gyration ($R_g$) of CoVE shows different non-monotonic dependence on $c$ (increase followed by decay) in native and denature phase with a higher and sharper rate of change in farmer. Effective dimension ($D$) of CoVE is estimated from the scaling of structure factor: in denatured phase, $D \approx 2$ (a random coil conformation) at low $c$ (= $0.01 - 0.10$) with appearance of some globularization i.e. $D \approx 2.3, 2.5$ at higher $c$ (= $0.2, 0.3$). Increasing $c$ seems to reduce the globularity ($D \approx 3$) of CoVE in native phase.

**Introduction**

Under the superb guidance of Dietrich Stauffer [1], I started my computer simulations to investigate random walk through percolating system [2] over three decades ago [3]. Anomalous diffusion [4] on a percolating network, a heterogeneous fractal, was one of our findings that led me to understand and appreciate the importance of multiscale structure and dynamics in addressing diverse issues throughout my research career. Dietrich Stauffer became a life-long collaborator, mentor, and friend ever since. I was very fortunate to work with a prolific scientist and unconventional educator and human being [1], I cannot express my gratitude enough to Dietrich Stauffer for a life-long education.

In this article, I will focus on the stochastic movements of a novel corona virus envelop protein, CoVE, [5-8] in porous media generated by a random distribution of impenetrable quenched barriers [2]. Random walk motion of a particle through percolating systems particularly on a cubic lattice has been extensively studied by computer simulations for decades. The global transport of a particle hopping from site-to-site on conducting/pore sites occurs only at and above the percolation threshold and ceases below [2-4]. It is now well-known [4] that the random walk motion is diffusive above the percolation threshold and anomalous at the threshold in asymptotic time regime.

Percolating transport for the stochastic motion of such complex constituent as a protein is expected to be very different from that of a particle. Unlike the conventional 'transport of a particle through percolating porous media'[2-4], the fraction/concentration of forbidden sites occupied by impenetrable barriers will be relatively small in this work (as we see below), due to complex nature of the protein with a fluctuating shape. Note that the porosity must be at or above the percolation threshold (about 0.312 for a simple cubic lattice [2]) for an asymptotic global transport of a particle across the sample [2-4]. The threshold porosity for the transport of a protein chain would be much higher than that for the conventional transport of a particle. To my knowledge, this is the first attempt to study the stochastic motion of a CoVE protein chain through such a random porous medium.

In addition, probing the structural relaxation of a small protein, CoVE, [8-16] of corona virus in heterogeneous media would be interesting in context of current COVID-19 pandemic. The envelope protein (CoVE) is believed to be a transmembrane ion-channel protein and plays an important role in morphogenesis of the corona virus and its assembly. Membrane proteins in ion channels are generally in heterogeneous matrix and play a key role in creating a specific pathway for selective transport of various critical elements such as ions, water, etc. for cell functions. Despite enormous efforts, detailed structural information on CoVE proteins is very limited. A random distribution of impenetrable barriers [2] is used here to create a porous medium that constitutes a simple host matrix in which a CoVE protein is embedded. Native and denatured phases of protein structures are controlled by temperature (typically low and high respectively). We have recently examined the structure and dynamics of CoVE as a function of temperature [17] and found that the size of the protein depends non-monotonically on temperature over the entire native to denature phase. In this article, we explore the structure and dynamics of CoVE protein in porous media at a low and a high temperature in native and denature phase respectively.

'Protein folding' has been extensively studied by computer simulations for over half a century and remains an open problem still [18, 19]. A large fraction of these investigations relies on various molecular model (with atomistic details), speculative interpretation of laboratory observations, and ad hoc assumptions. In order to develop a sound understanding it is useful to examine the structure and dynamics of such constitutive proteins based on the fundamental principles of statistical physics. The structure and dynamics of a protein depends on temperature, solvent, and a number of factors of the underlying host environment in which the protein performs its stochastic movements. Time scale with all-atom approaches is a major bottleneck due to huge degrees of freedom. Coarse-graining provides a practical approach to probe various issues in order to gain some insight. There are numerous coarse-grained methods [20-32] used in such studies. We have been using a coarse-grain approach to examine structure and dynamics of a number of proteins over the years [33-38]. It involves (*i*) a bond-fluctuation mechanism used extensively in modeling polymer on a cubic lattice with ample degrees of freedom [39], a lesson learnt from computational polymer and (*ii*) knowledge-based residue-residue interactions used extensively in protein folding to capture specificity [40-45]. This approach described in brief in the following is implemented here to study structure and dynamics of CoVE in porous media.

**Model**

We consider a coarse-grained model of protein on a cubic lattice where a residue is represented by a cubic node of size $(2a)^3$ where $a$ is the lattice constant [33-38]. CoVE protein consists of 76 residues in a specific sequence [10] and is represented by *76* nodes tethered together by flexible covalent bonds; the bond length between consecutive nodes varies between *2* and $\sqrt{(10)}$ in unit of the lattice constant [39]. The specificity of a residue is considered by its unique interaction (see below). The protein chain is initially placed in a random configuration by implementing the excluded volume constraints. The porous medium is generated by distributing impenetrable immobile barriers randomly with excluded volume conditions in presence of the protein chain on a fraction *c* of the lattice sites. A barrier is represented by a cube of the same size as that of the node. Each residue interacts with surrounding residues within a range ($r_c$) with a generalized Lennard-Jones potential,

$$U_{ij} = \left[ \left| \varepsilon_{ij} \right| \left( \frac{\sigma}{r_{ij}} \right)^{12} + \varepsilon_{ij} \left( \frac{\sigma}{r_{ij}} \right)^{6} \right], \; r_{ij} < r_c \quad (1)$$

where $r_{ij}$ is the distance between the residues at site $i$ and $j$; $r_c= \sqrt{8}$ and $\sigma = 1$ in units of lattice constant. Only excluded volume interaction between the barrier and residues is considered. A knowledge-based residue-residue interaction matrix [40] is used as input for the potential strength $\varepsilon_{ij}$. The knowledge-based residue-residue contact matrix has been developed and extensively used over the years. It is derived from a large (and growing) ensemble of protein structures in protein data bank (PDB) [45]. The residue-residue contact matrix elements consist of positive and negative values with appropriate magnitudes that incorporates the hydrophobic, polar, and electrostatic characteristics of the amino acids. Estimates of the contact matrix elements involve approximations and assumptions [40-45] as usual in statistical analysis of the residue contacts. The generalized LJ potential (1), based on the knowledge-based interaction is thus phenomenological.

Each residue performs its stochastic movement with the Metropolis algorithm, i.e. with the Boltzmann probability $exp(-\Delta E/T)$ where $\Delta E$ is the change in energy between new and old positions of the residue and $T$ is the temperature in reduced unit of the Boltzmann constant. Unit Monte Carlo step is defined by attempts to move each residue once. Both local and global physical quantities such as radius of gyration, structure factor, contact map, mobility profile, etc. are examined some of which are presented here. Most of our simulations are performed on a $150^3$ cubic lattice with barrier concentration $c = 0.01 – 0.30$ at a low temperature $T = 0.020$ (native phase) and a high temperature $T = 0.030$ (denatured phase) [17]. Simulations are also carried out with different lattice sizes to test for the abnormal effects of the finite size on the qualitative nature of observations. The number of independent samples $10–500$ varies, lower number of samples with long simulations and higher barrier concentrations and larger number of samples with shorter runs and low barrier concentrations.

**Results and discussion**

Structural response of a CoVE protein chain in a simulation box without barriers (i.e. $c=0$) is already studied [17] as a function of temperature which exhibits a non-monotonic thermal response. The conformation of the protein undergoes a continuous transition [17] with temperature, from a globular structure in native phase ($T \leq 0.02$) to a random coil conformation in denatured phase ($T=0.03$). The structure of protein in both native and denature phase is affected by the presence of barrier and depends on the barrier concentration $c$. The dynamics of protein is very slow (see below) in native phase in comparison to that in denatured phase. Most data presented in the main text is focused on results of simulations in denatured phase ($T=0.03$); corresponding results, as appropriate, in native phase ($T=0.02$) are included in the supplement.

Figure 1 shows illustrative snapshots of the protein in denatured phase ($T=0.03$) with barrier concentrations $c = 0.04$ and $0.10$. Obviously, the protein is more confined at higher barrier concentration; movements of protein and its conformational relaxation are accordingly more restricted at higher barrier concentration. Relaxation of protein conformation is enhanced with reduced confinement at low barrier concentration $c = 0.04$ which leads to a more compact conformation than that at $c = 0.10$ where the protein is essentially trapped. Corresponding snapshots of the protein conformations in native phase ($T=0.02$) shows the effect of confinement on the conformational relaxation of the protein (see figure S1). Since the protein is globular in native phase, it is relatively more compact at low barrier concentration (*c = 0.04*) in a relatively lower confinement than that at the higher barrier concentration (*c = 0.10*). Apart from the difference in the conformations of the protein in its native and denature phases, the effect of barrier on the conformational relaxation is qualitatively similar with regard to confinement.

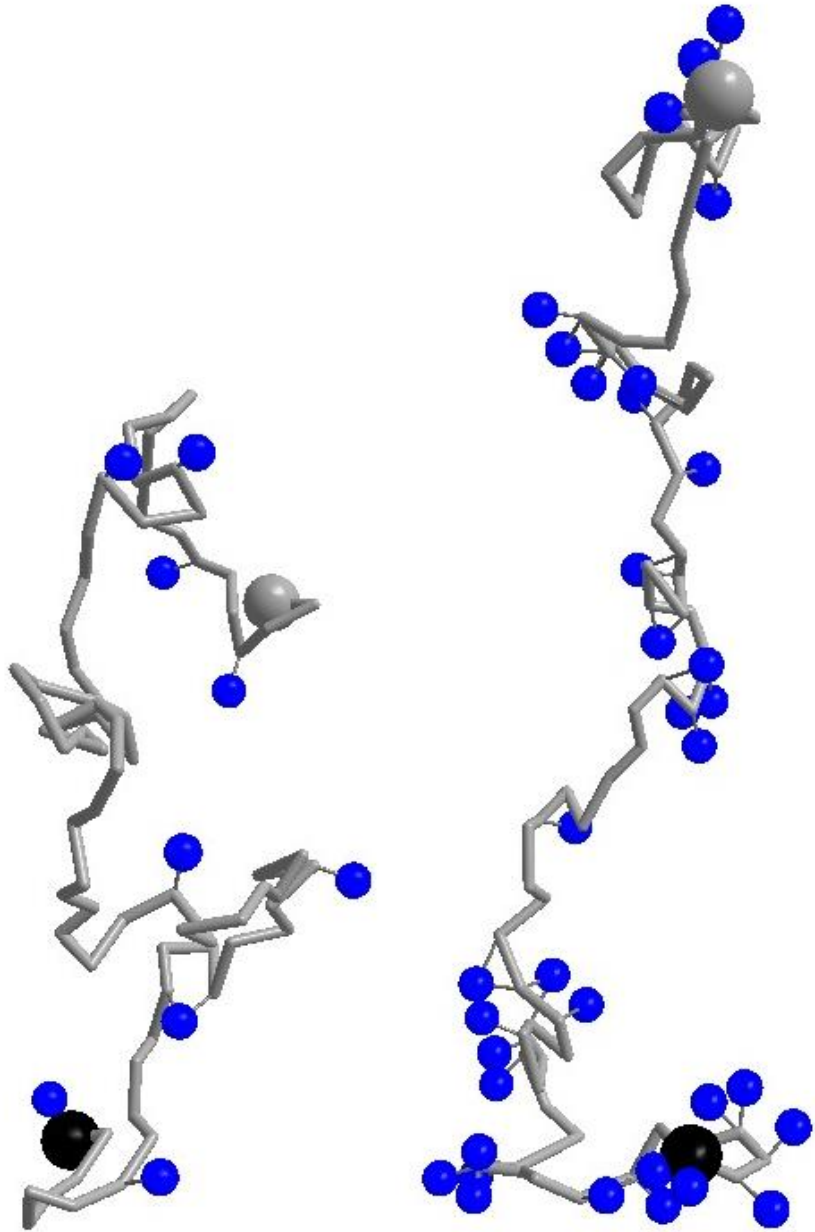

Figure 1: Snapshot of CoVE protein backbone at the end of $10^7$ MCS time in denatured phase (*T=0.030*) at barrier concentrations *c = 0.04* (left) and *c = 0.10* (right). Large black sphere represents the first residue ($^1$M) and grey, the last ($^{76}$V) of the protein. Smaller spheres represent the barrier particles within the range of interaction of residues (connected thin lines).

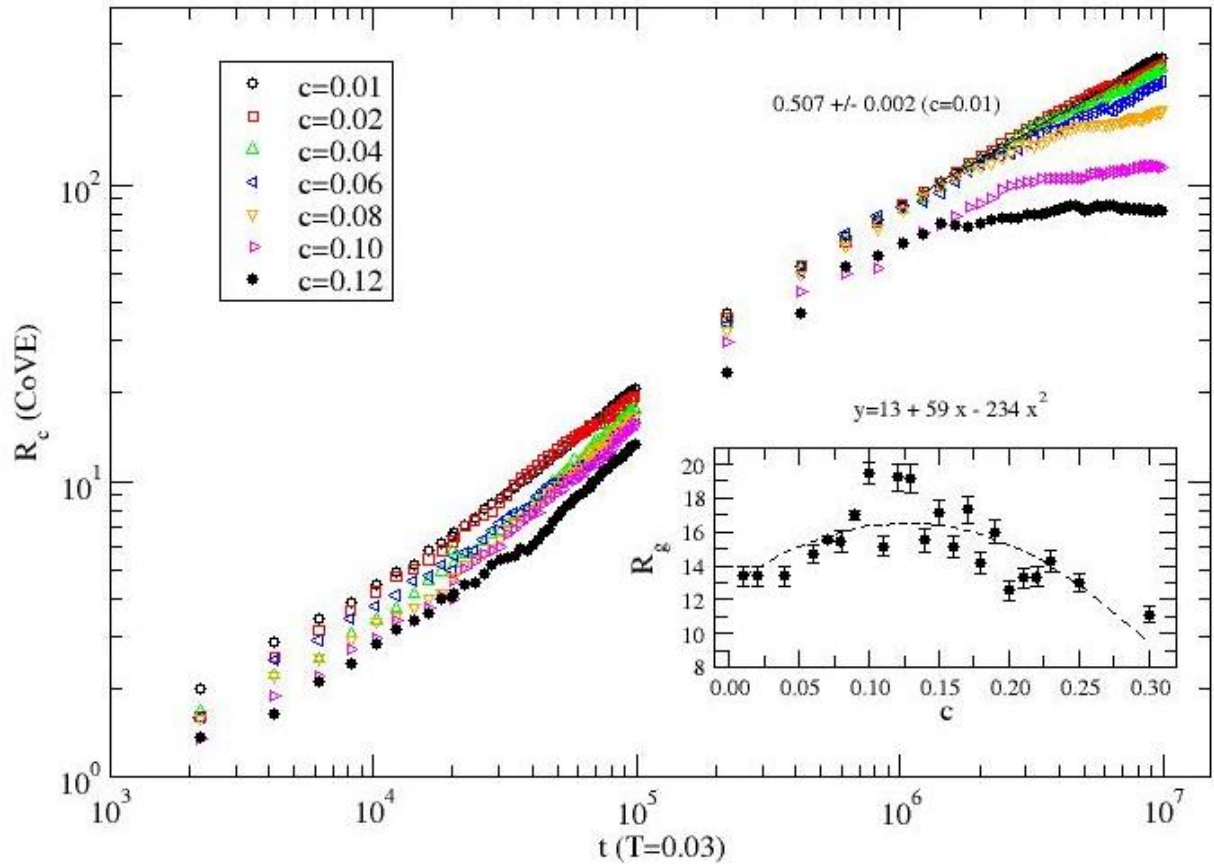

Figure 2: Variation of the root mean square displacement ($R_c$) of CoVE with the time step ($t$) on a log-log scale for a range of barrier concentration $c = 0.01 - 0.12$ in denature phase ($T=0.030$). The slope of the asymptotic data is included (fitted line in black). Inset is the variation of the radius of gyration of the protein with the barrier concentration; a polynomial (included on top) fit of data is shown by broken line.

Variations of the root mean square displacement (RMSD) of the center of mass ($R_c$), with time step ($t$) is presented in figure 2 as a function of barrier concentration ($c$) at the temperature $T = 0.03$. The nature of the asymptotic variation of $R_c$ with $t$ clearly shows how the long time global dynamics of protein depends on $c$. Analysis of a power-law dependence, $R_c \propto t^\nu$, can help characterizing the type of dynamics, i.e. diffusive if $\nu = \frac{1}{2}$. At low barrier concentrations (i.e. $c = 0.01, 0.02$), the dynamics of protein is diffusive. It begins to deviate with lower value of the power-law exponent $\nu$ on increasing the barrier concentration i.e. $c = 0.08$. The global motion ceases with $\nu = 0$ at $c \geq 0.10$. The threshold porosity $c_p = 1-c = 0.9$ for the percolation of a denatured CoVE protein is much higher than the percolation threshold ($\approx 0.312$) of a site percolation [2] which is the minimum fraction of open sites for a particle to continue its global transport. The effective threshold, thus, depends on the constitutive element and is very high for the protein to continue its global motion.

A similar analysis of the RMSD of the center of mass in native phase ($T=0.02$) shows that the effect of barrier on the dynamics of protein in native phase (figure S2) is different from that in denatured phase. The protein moves very slowly (figure S2) even at a low barrier concentration ($c = 0.01$). However, it begins to move faster, in contrast to denature phase, on increasing the barrier concentration ($c =0.10-0.20$) – barrier-induced motion. The global motion of the protein eventually stops at high barrier concentration ($c \geq 0.30$). Since the residue-residue interaction (attractive overall) dominates over the thermal energy in native phase ($T=0.02$), the asymptotic dynamics becomes extremely slow as the protein relaxes into a compact configuration. Presence of barriers at a moderate concentration (i.e. $c =0.10-0.20$) prevents some segmental globularization resulting in a reduced residue interaction energy. The stochastic movements of some residues are enhanced due to thermal noise that leads to a sub-diffusive (figure S2) global dynamics. At high barrier concentration ($c =0.30$), the confinement is enhanced so much that the stochastic movement of majority of residues ceases to occur and the protein is trapped.

The radius of gyration ($R_g$) of the protein depends on the barrier concentration. Because of large fluctuations in data (see the inset in figure 2) it is rather difficult to identify a clear trend. However, it appears to depend non-monotonically on the barrier concentration i.e. increase of $R_g$ followed by decay with $c$. In dilute regime ($c =0.01-0.04$), $R_g$ remains constant, it increases with the concentration in the range $c =0.06-0.12$, then fluctuate wildly (samples are not good enough to identify the trend) before it decays in the concentration range $c =0.24-0.30$. Increase in $R_g$ appears due to pinning of some residues as others perform its stochastic movements. Decay in $R_g$ occurs due to enhanced confinement of residues with increased number of surrounding barriers (see figure 1). On increasing the barrier concentration in native phase ($T=0.02$) (see the inset in figure S2), the radius of gyration of the protein remains almost constant at low concentration ($c =0.01-0.04$), followed by a continuous increase (i.e. $c = 0.10-0.20$) until a certain concentration (i.e. around $c =0.23$) beyond which it decays to a constant where the conformation is fully arrested within the confinement of the barriers.

Let us examine the contact profiles of residues with barriers and residues. The average number $N_c$ of barriers around each residue can be a measure of confinement while the average number $N_n$ of residues around each residues provides an average contact map of its segments. Figure 3 shows the confinement map of the protein at the barrier concentration $c = 0.01-0.10$ at $T = 0.03$. The profile is relatively uniform along the protein contour at each barrier concentration, i.e. the magnitude of $N_c$ remain almost the same for each residue. Obviously, there is almost no confinement at $c = 0.01$ with negligible value of $N_c$ as one would expect at such a low barrier concentration. The magnitude of $N_c$ increases with the barrier concentration. With an average value of $N_c \geq 1$ at the barrier concentration $c \geq 0.10$ the asymptotic global motion of the protein chain ceases (see figure 2). A similar trend in confinement map of the protein is also observed in its native phase ($T=0.02$) with a somewhat more fluctuating profile (see figure S3). Since the protein conform to a globular conformation at $T=0.02$, some residues are confined by the presence of surrounding residues itself. Therefore, the probability of encounter with the barriers varies with residues (interior versus exterior) along the protein backbone in its native conformation ($T=0.02$) unlike in denatured phase ($T=0.03$). The residue-residue contact profiles ($N_n$) show negligible segmental coagulation at low barrier concentrations ($c = 0.01-0.10$) in both native ($T=0.02$) and denatured ($T=0.03$) phases (see figure S4, S5). However, in denature phase ($T=0.03$) contacts among selected residues become relatively higher at high barrier concentrations, more at $c = 0.20$ than that at $c = 0.30$ (see figure S6), barrier induced segmental globularization. Thus, depending on concentration, presence of barriers inhibits the self-assembly of residues as well as enhances their contacts via confinement.

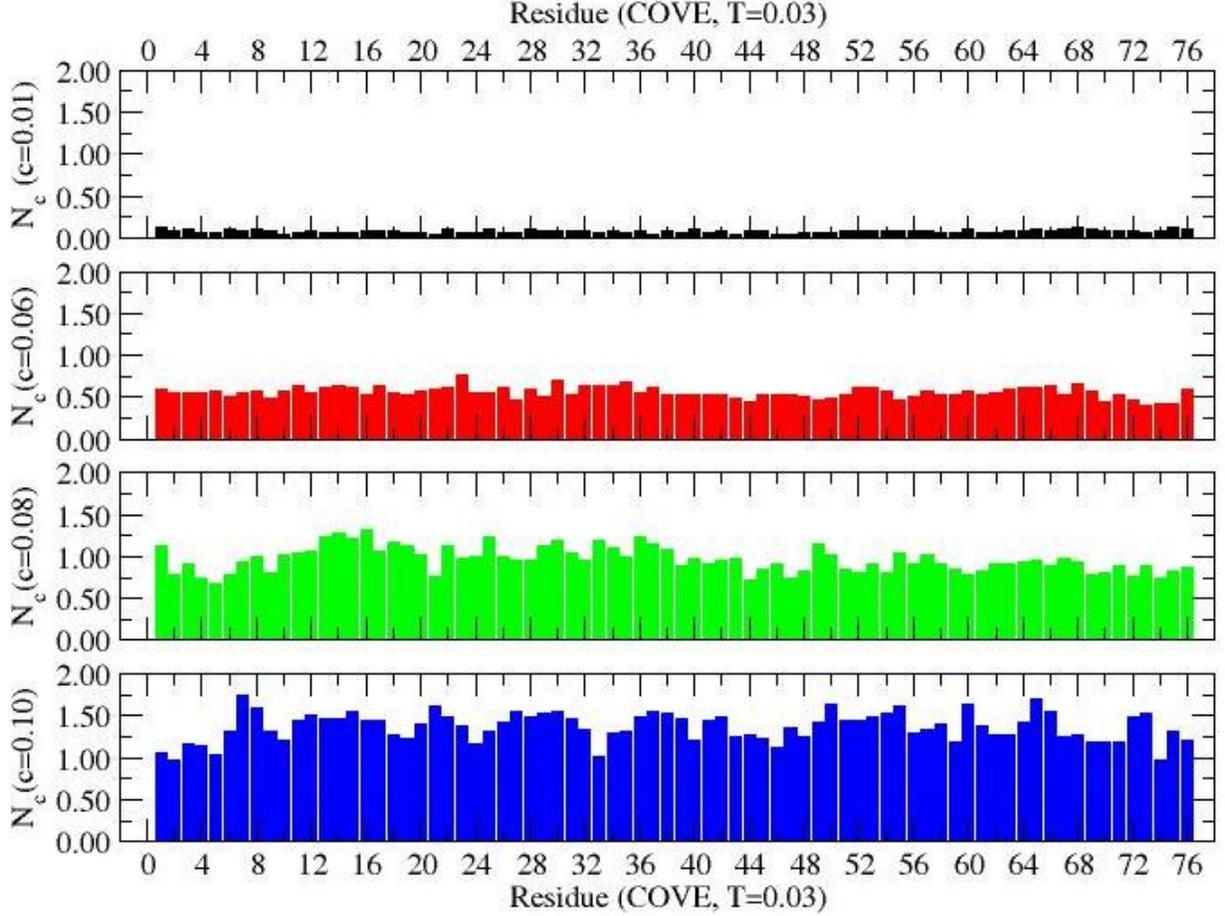

Figure 3: Contact profile of residues with barriers at concentration $c = 0.01$, $0.06$, $0.08$, and $0.10$ in denature phase ($T=0.030$); $N_c$ is the average number of barrier particles around each residue.

The conformation of the protein chain can also be analyzed by examining the structure factor

$$S(q) = \langle \frac{1}{N} \left| \sum_{j=1}^{N} e^{-i\vec{q} \cdot r_j} \right|^2 \rangle_{|\vec{q}|} \qquad (2)$$

where $r_j$ is the position of each residue and the magnitude of the wave vector ($q$) of wavelength $\lambda$ is given by $|q| = 2\pi/\lambda$. With a power-law scaling for the structure factor, $S(q) \propto q^{-1/\gamma}$, one may be able to evaluate the power-law exponent $\gamma$. Therefore, from scaling of the radius of gyration ($R_g$) of the protein chain (comparable with the wavelength $\lambda$) with the number of residues ($N$), i.e. $R_g \propto N^\gamma$, one can estimate the the effective dimension $D = 1/\gamma$ [33-38]. Figure 4 shows the variations of $S(q)$ with the wavelength $\lambda$ (lambda) comparable to radius of gyration of the protein (figure 2) at the barrier concentrations $c = 0.01 - 0.30$ in denatures phase ($T=0.03$). The effective dimension $D \approx 2$ at lower barrier concentrations ($c = 0.01 - 0.10$) shows that the protein remains in a random coil conformation (see figure S7). Enhanced confinement at high barrier concentration thus leads to increasing globularity with $D \approx 2.3, 2.5$ ($c = 0.2, 0.3$) (figure 4) in denatured phase ($T=0.03$). In a similar analysis of the structure factor in native phase (figure S8), the conformation of the protein is found to be globular ($D \approx 3$) at low $c$. The globularity is reduced somewhat (*i.e.* $D \approx 2.7$) in native phase (figure S8) at high barrier concentrations (*i.e.* $c = 0.19, 0.23$).

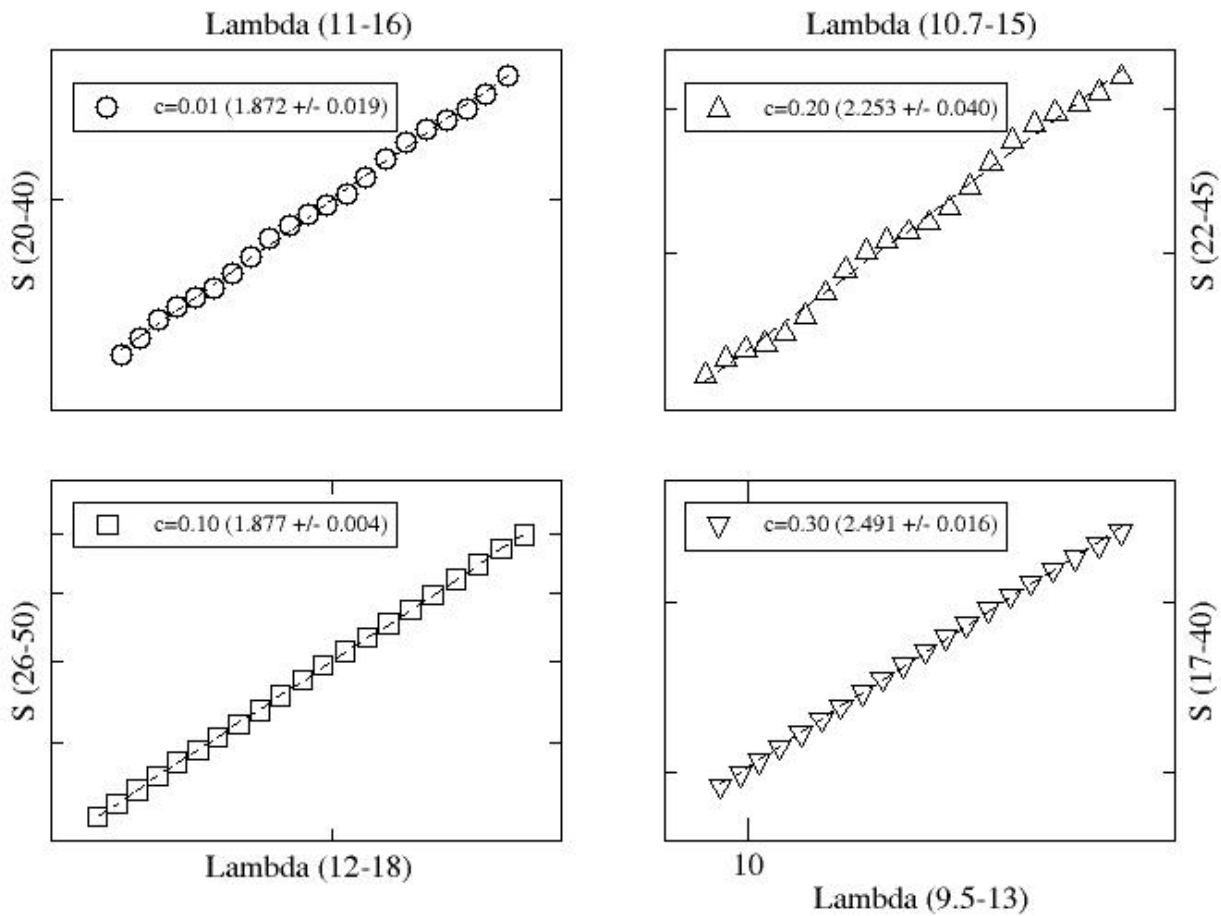

Figure 4: Structure factor (S) of the protein versus wave length (lambda) at representative barrier concentrations *c =0.01*, *0.10*, *0.20*, and *0.30* in denature phase (*T=0.030*) with the estimates of slopes. Data points are selected with wavelength comparable to radius of gyration of the protein at corresponding concentrations. The number along the axes are the range of the axes (starting and end points) for a guide.

**Conclusion**

A Monte Carlo simulation is used to investigate structure and dynamics of a protein (CoVE) in random media generated by a random distribution of barriers at concentration *c* with a coarse-grained model. Large-scale simulations are performed for a range of barrier concentration (*c*) in native (*T=0.02*) and denature (*T=0.03*) phase of the protein. In denature phase, protein is diffusive at low barrier concentrations (*c = 0.01, 0.02*) and slows down on increasing the barrier concentration before its global motion stops at $c \geq 0.10$. The threshold porosity ($c_p = 1-c$) for the percolation of the CoVE protein in denatured phase is found to be around 0.9. Note that $c_p$ is much higher than the percolation threshold ($\approx 0.312$) of a site percolation which is the minimum fraction of open sites for a particle to continue its global transport. The effective threshold ($c_p$) depends on the percolating element and very high for a protein (CoVE) with fluctuating shapes to continue its global motion.

The effect of barriers on the dynamics of protein in native phase is different from that in denatured phase. The protein moves faster on increasing the barrier concentration in a characteristic range (*c =0.10–0.20*) – barrier-induced dynamics but eventually stops at high barrier concentration ($c = 0.30$). Obviously, confinement is enhanced at high barrier concentration ($c = 0.30$) so much that the stochastic movement of majority of residues ceases to occur and the protein is trapped. The threshold porosity for CoVE protein in its native phase ($c_p \sim 0.7$) to percolate is lower than that in its denatured phase. Concept of percolation for the transport of a particle executing its random walk through a percolating (conducting) porous medium is not the same as the transport of a object with fluctuating shape like the protein performing its stochastic movements. In the paragidm of generic percolation, Stauffer would have labelled the author a 'enemy/traitor' in his humerous comments.

In denature phase, the radius of gyration ($R_g$) of the protein appears to exhibit a non-monotonic dependence on the barrier concentration, i.e. increase of $R_g$ is followed by decay with the barrier concentration. The radius of gyration ($R_g$) of the protein in native phase increase with the concentration in a characteristic range ($c \approx 0.10 - 0.23$) before decaying sharply to a saturated value.

Contact profiles of the protein with barriers and residues are studied by calculating the average number of surrounding barriers ($N_c$) and residues ($N_n$) within the range of interactions of each residue. The barrier contact profiles of protein in denatured phase is almost uniform along its backbone at each barrier concentration. The value of barrier contact number over which the protein is trapped by confinement is found to be $N_c \geq 1$ at the barrier concentration $c \geq 0.10$. The barrier contact profile of the protein in its native phase fluctuates along the backbone. In general, the presence of barrier reduces the segmental globularization in both native and denature phase. However, in denature phase contacts among selected residues become relatively higher at high barrier concentrations (more at $c = 0.20$ than that at $c = 0.30$) - barrier induced segmental globularization.

Scaling analysis of the structure factor provides a quantitative estimate of its effective dimension ($D$), a measure of globularity of the protein conformation. The protein remains in a random coil conformation ($D \approx 2$) in denatured phase at low barrier concentration ($c = 0.01 - 0.10$). Confinement at high barrier concentration, however, leads to enhance globularity with $D \approx 2.3, 2.5$ ($c = 0.2, 0.3$) in denatured phase. In native phase, the protein remains globular specially at low $c$; the globularity is reduced slightly due to trapping at high $c$.

# Supplementary material

**A Monte Carlo simulation of a protein (CoVE) in a matrix of random barriers**

R.B. Pandey
School of Mathematics and Natural Sciences, University of Southern Mississippi, Hattiesburg, MS 39406-5043, USA
Email: ras.pandey@usm.edu

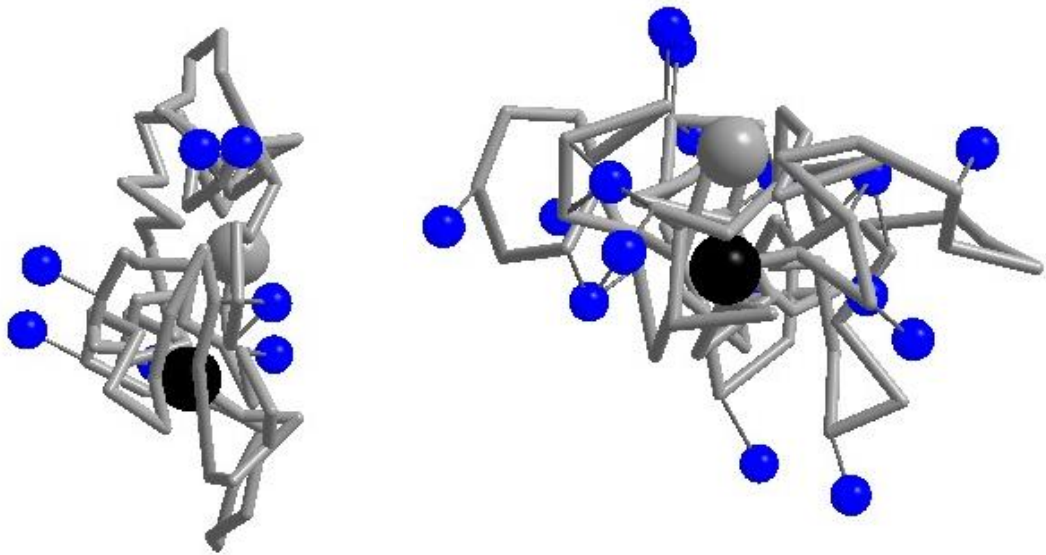

Figure S1: Snapshot of CoVE protein backbone at the end of $10^7$ MCS time in native phase (*T=0.020*) at barrier concentrations *c = 0.04* (left) and *c = 0.10* (right). Large black sphere represents the first residue ($^1$M) and grey, the last ($^{76}$V) of the protein. Smaller spheres represent the barrier particles within the range of interaction of residues connected by thin grey lines.

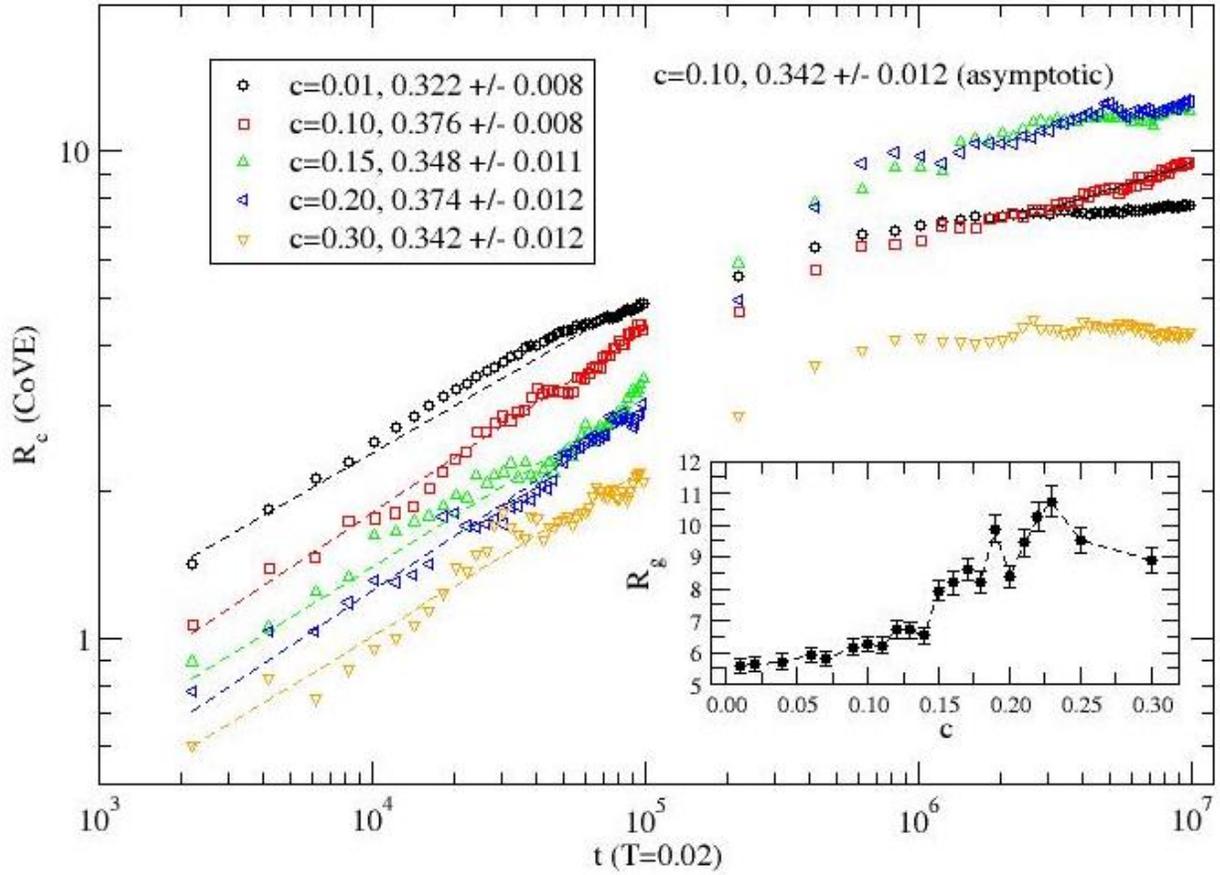

Figure S2: Variation of the root mean square displacement ($R_c$) of CoVE with the time step ($t$) on a log-log scale for a range of barrier concentration $c = 0.01 - 0.30$ in native phase ($T=0.020$). The slope of the asymptotic data for $c = 0.10$ is included (fitted line in black); slopes of data in short time ($t \sim 10^3 - 10^5$) regime is also included at each barrier concentration. Inset is the variation of the radius of gyration of the protein with the barrier concentration.

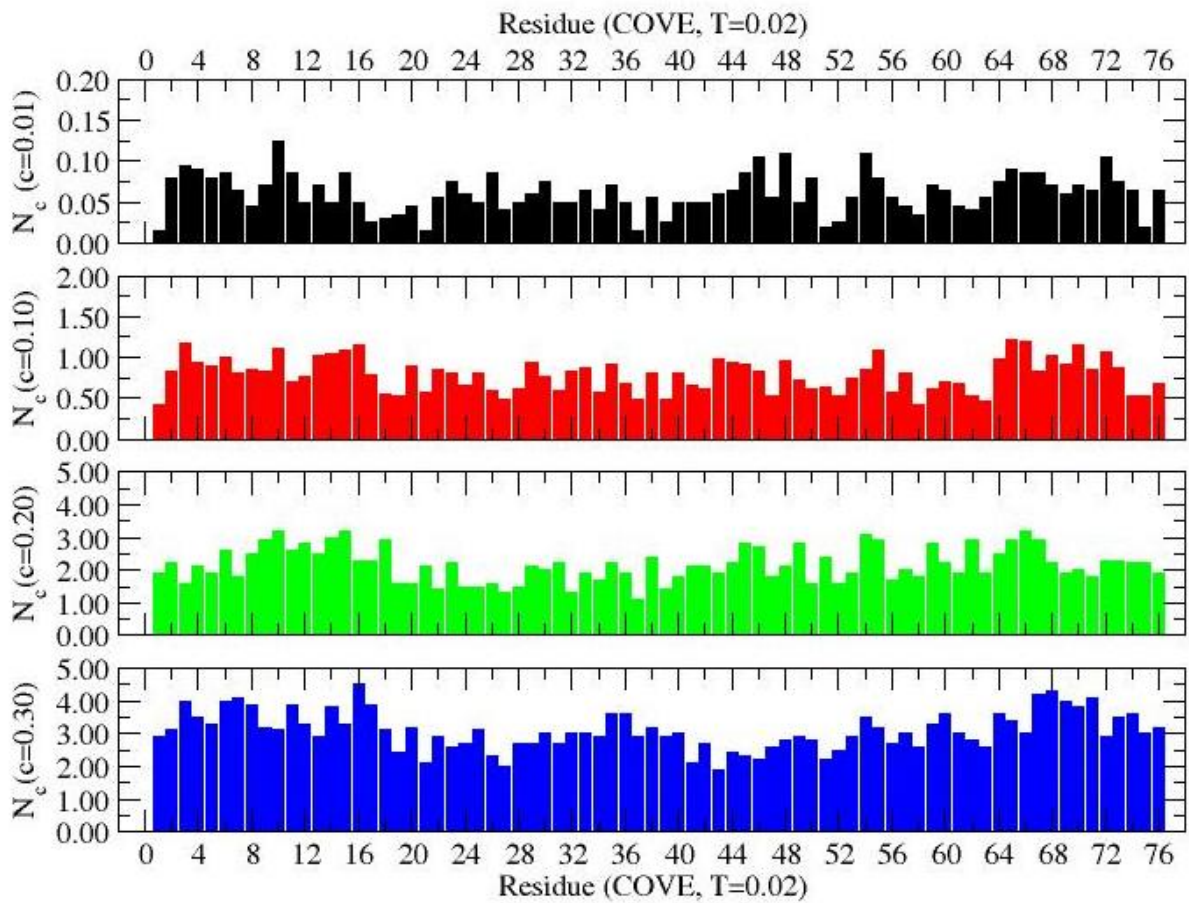

Figure S3: Contact profile of residues with barriers at concentration $c = 0.01$, $0.10$, $0.20$, and $0.30$ in native phase ($T=0.020$); $N_c$ is the average number of barrier particles around each residue.

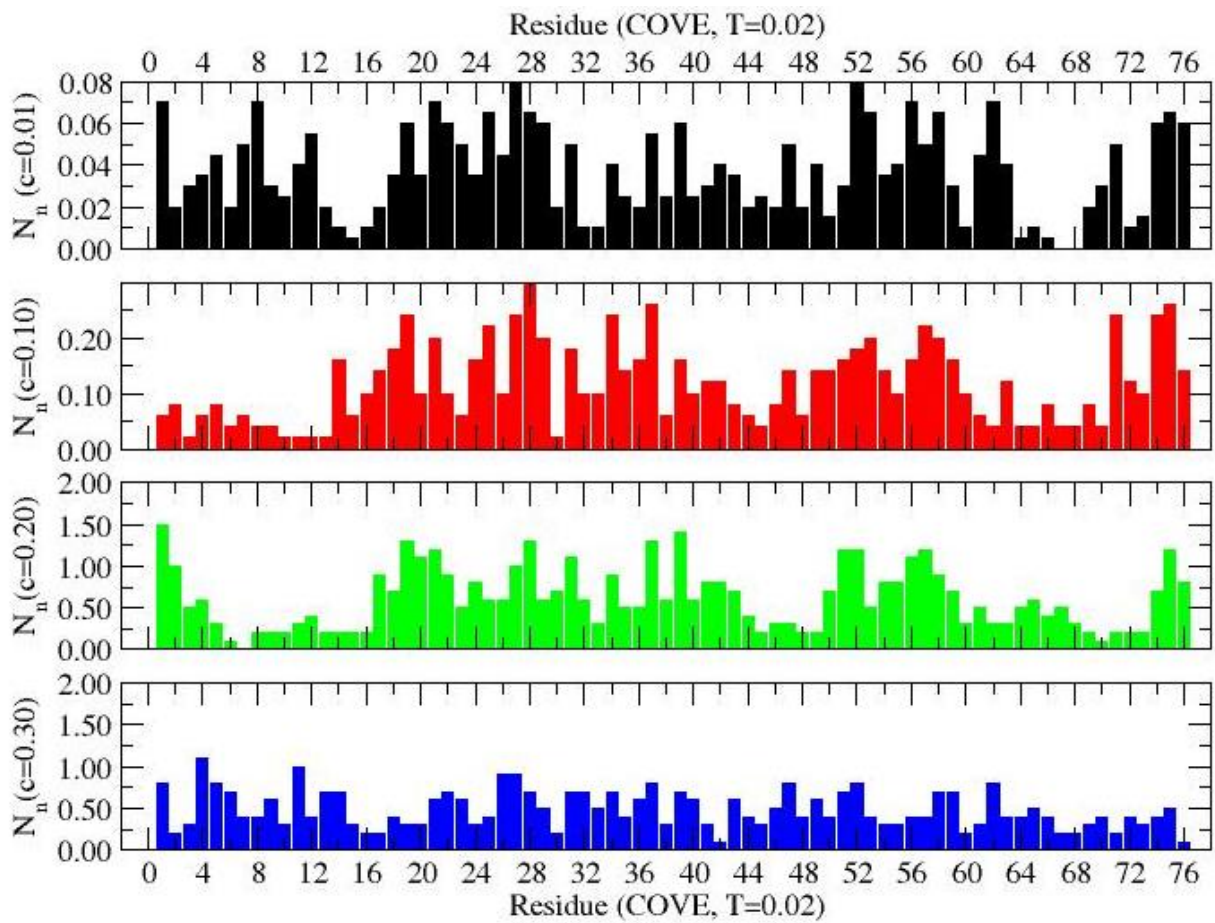

Figure S4: Contact profile of residues at concentration $c = 0.01$, $0.10$, $0.20$, and $0.30$ in native phase ($T=0.020$); $N_n$ is the average number of residues (excluding consecutively connected residues) around each.

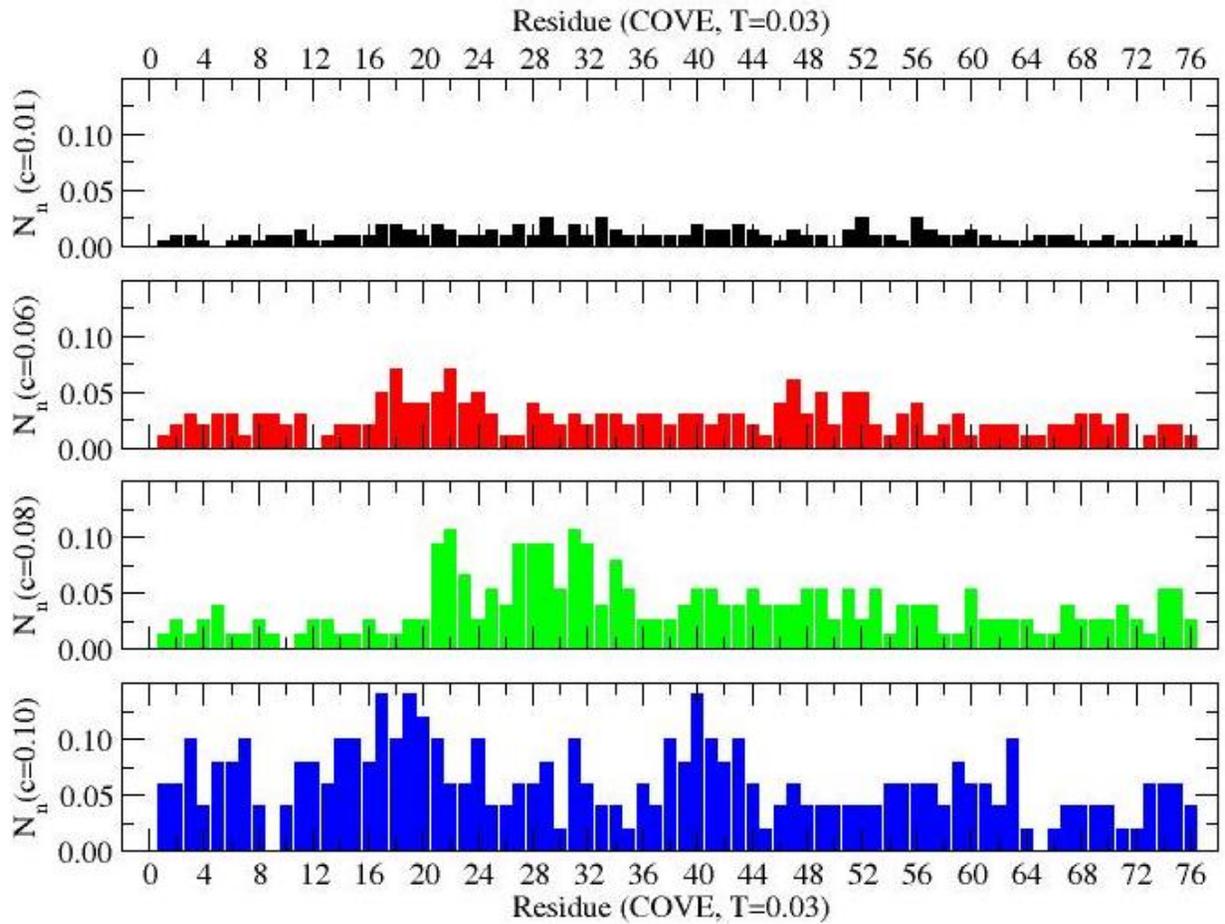

Figure S5: Contact profile of residues at concentration $c = 0.01$, $0.06$, $0.08$, and $0.10$ in denature phase ($T=0.030$); $N_n$ is the average number of residues (excluding consecutively connected residues) around each.

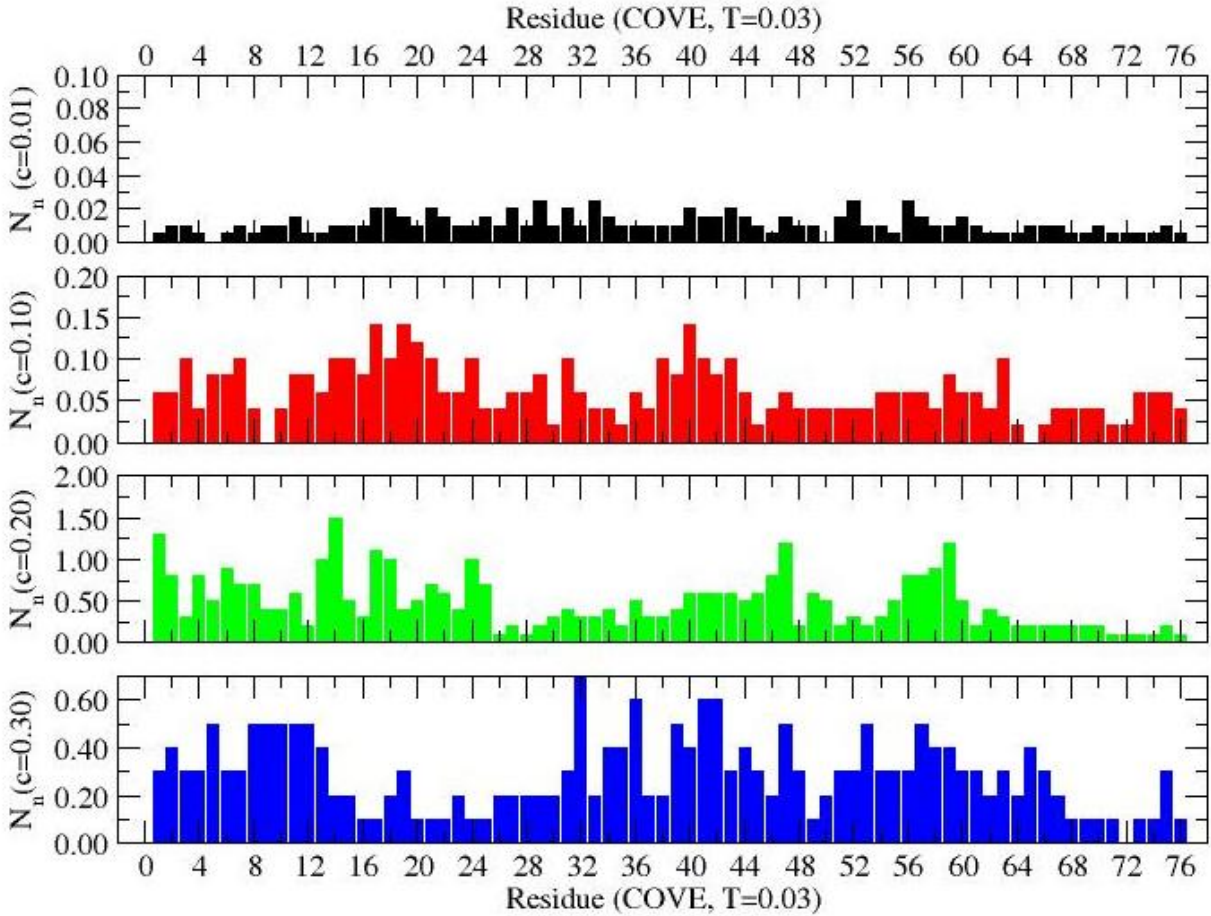

Figure S6: Contact profile of residues at concentration $c = 0.01$, $0.10$, $0.20$, and $0.30$ in denature phase ($T=0.030$); $N_n$ is the average number of residues (excluding consecutively connected residues) around each.

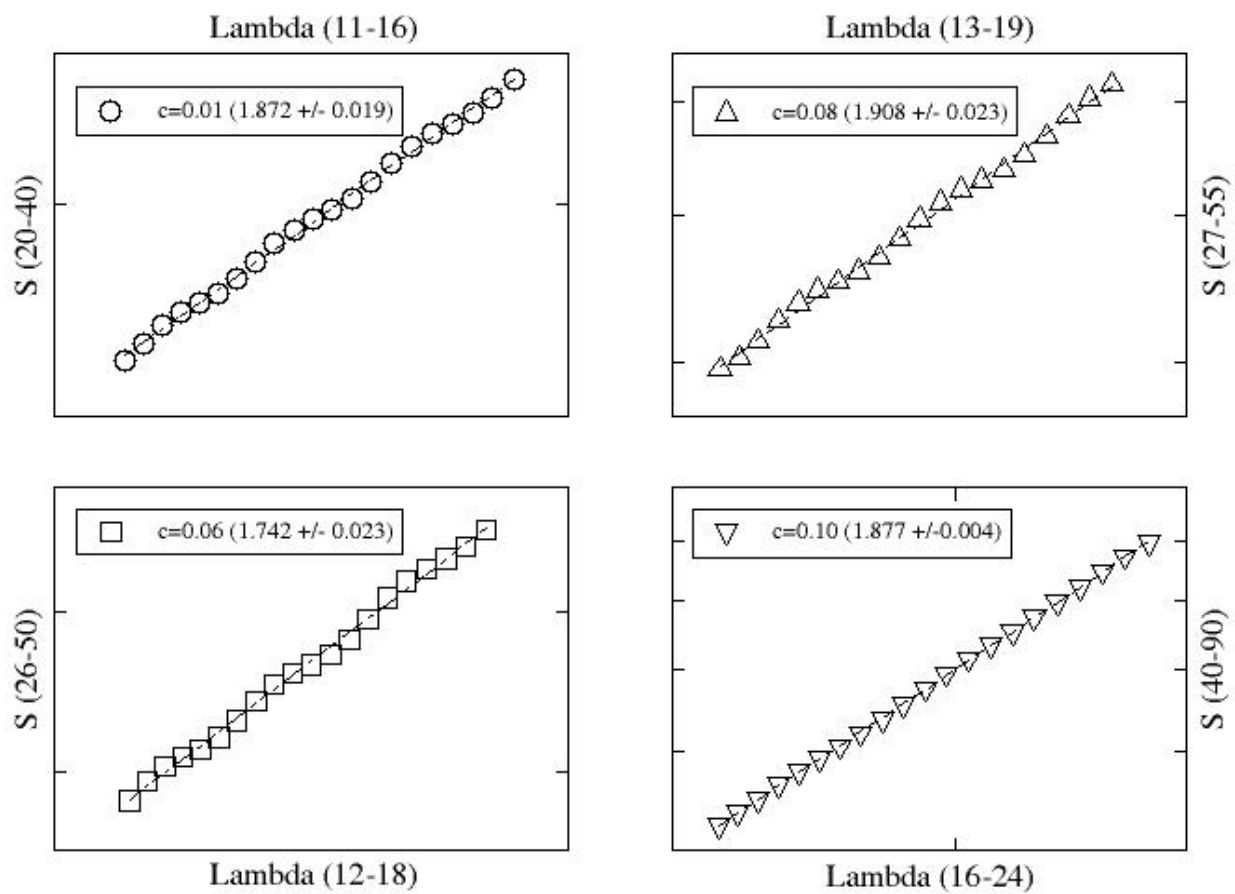

Figure S7: Structure factor (S) of the protein versus wave length (lambda) at representative barrier concentrations $c = 0.01$, $0.06$, $0.08$, and $0.10$ in denature phase ($T=0.030$) with the estimates of slopes. Data points are selected with wavelength comparable to radius of gyration of the protein at corresponding concentrations. The number along the axes are the range of the axes (starting and end points) for a guide.

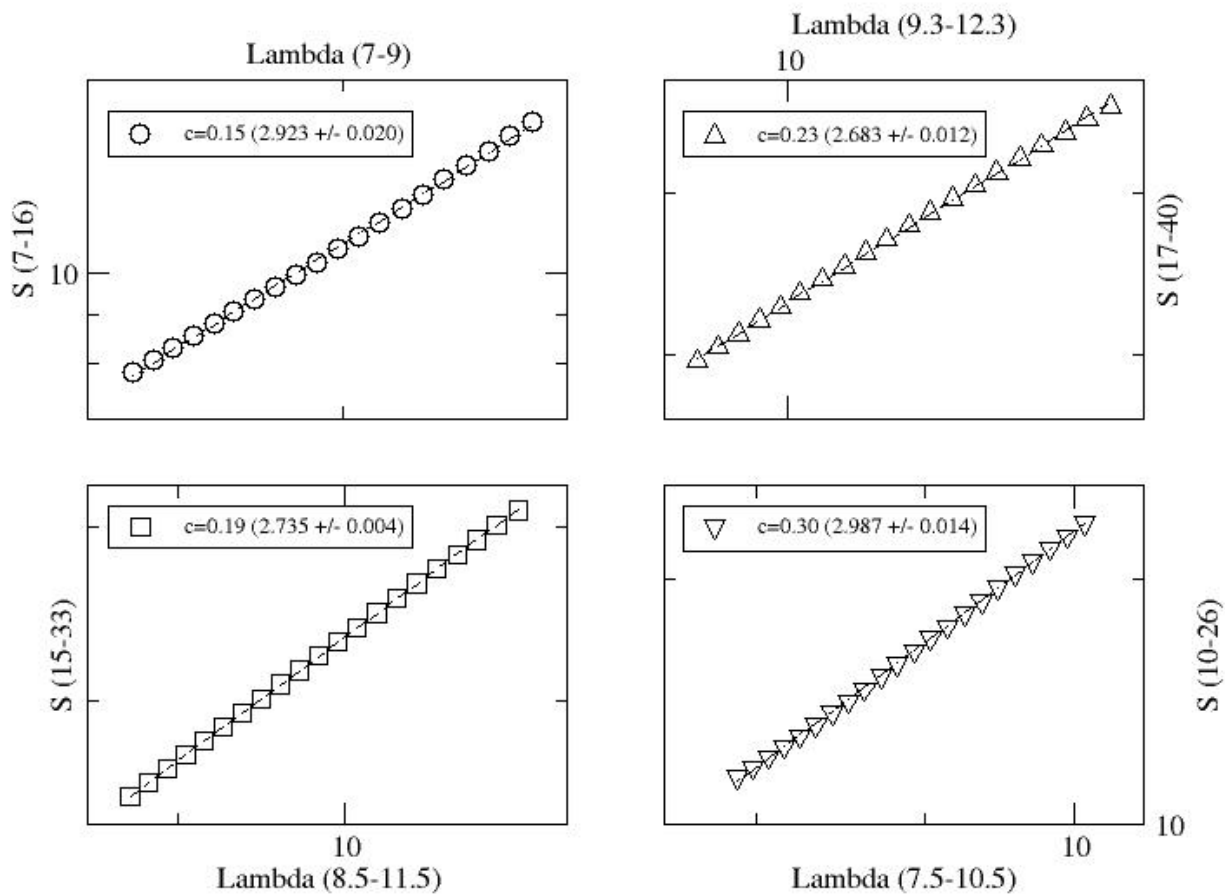

Figure S8: Structure factor (S) of the protein versus wave length (lambda) at representative barrier concentrations $c = 0.15$, $0.19$, $0.23$, and $0.30$ in native phase ($T=0.020$) with the estimates of slopes. Data points are selected with wavelength comparable to radius of gyration of the protein at corresponding concentrations. The number along the axes are the range of the axes (starting and end points) for a guide.